\documentclass[aps,prx,twocolumn,superscriptaddress]{revtex4-2}
\usepackage{graphicx}
\usepackage{latexsym}
\usepackage{amssymb}
\usepackage{amsmath}
\usepackage{amsfonts}
\usepackage{upgreek}
\usepackage{bm,dsfont}
\usepackage{multirow}
\usepackage{color}
\usepackage{hyperref}
\usepackage{ulem}
\hypersetup{
colorlinks = true,
linkcolor = [rgb]{0.70,0.13,0.13},
citecolor = [rgb]{0.13,0.55,0.13},
urlcolor = [rgb]{0.25, 0.41, 0.88}}

\newcommand{\dd}{\mathrm{d}}
\newcommand{\ii}{\mathrm{i}}

\newcommand{\U}{\mathrm{U}}
\newcommand{\SU}{\mathrm{SU}}

\newcommand{\SO}{\mathrm{SO}}

\newcommand{\scK}{\mathcal{K}}

\newcommand{\vect}[1]{{\bm{#1}}}

\newcommand{\mat}[1]{\left[\begin{matrix}#1\end{matrix}\right]}

\newcommand{\eq}[1]{\begin{equation}#1\end{equation}}
\newcommand{\eqs}[1]{\begin{equation}\begin{split}#1\end{split}\end{equation}}
\newcommand{\eqnref}[1]{Eq.\,\eqref{#1}}
\newcommand{\figref}[1]{Fig.\,\ref{#1}}
\newcommand{\tabref}[1]{Tab.\,\ref{#1}}

\newcommand{\appref}[1]{Appx.\,\ref{#1}}
\newcommand{\refcite}[1]{Ref.\,\onlinecite{#1}}

\begin{document}

\title{ Kohn-Luttinger Superconductivity and Inter-Valley Coherence\\ in Rhombohedral Trilayer Graphene}

\author{Yi-Zhuang You}
\affiliation{Department of Physics, University of California, San Diego, CA 92093, USA}


\author{Ashvin Vishwanath}
\affiliation{Department of Physics, Harvard University, Cambridge, MA 02138, USA}

\date{\today}
\begin{abstract}

Motivated by recent experiments on ABC-stacked rhombohedral trilayer graphene (RTG) which observed spin-valley symmetry-breaking and superconductivity, we study instabilities of the RTG metallic state to symmetry breaking orders. We find that interactions select the inter-valley coherent order (IVC)  as the preferred ordering channel over a wide range, whose theoretically determined phase boundaries agree well with experiments on both the hole and electron doped sides. The Fermi surfaces near van Hove singularities admit partial nesting between valleys, which promotes  both inter-valley superconductivity and IVC fluctuations. We investigate the interplay between these fluctuations and the Hunds (intervalley spin) interaction using a renormalization group approach. For antiferromagnetic Hund's coupling,  intervalley pairing appears in the spin-singlet channel with enhanced $T_c$, that scales with the dimensionless coupling $g$ as $T_c\sim\exp(-1/\sqrt{g})$ , compared to the standard $\exp(-1/g)$ scaling. In its simplest form, this scenario assumes a sign change in the Hund's coupling on increasing hole doping. On the other hand, the calculation incorporates  breaking of the independent spin rotations between valleys from the start, and strongly selects spin singlet over spin triplet pairing, and naturally occurs in proximity  to the IVC, consistent with observations.  
\end{abstract}
\maketitle

{\it Introduction.}~--- The recent discovery of spin-valley-ordered metals\cite{Zhou2021Half} and superconductivity\cite{Zhou2021Superconductivity} in the ABC-stacked rhombohedral trilayer graphene (RTG) provides a new platform to investigate strongly correlated phases of matter in graphene-based systems\cite{Feldman2009Broken-symmetry,Mayorov2011Interaction-Driven,Cao2018,Cao2018a,Yankowitz2018Tu,Yankowitz2019Tuning,Lu2019Superconductors,Chen2019Signatures,Liu2020Tunable,Singh-Arora2020Superconductivity,Andrei2020Graphene,Balents2020Superconductivity,Hao2021Electric,Park2021Tunable}. The RTG band structure features van Hove singularities with the divergent density of states (DOS) near charge neutrality\cite{Zhang2010Band}, which can be further enhanced by a perpendicular electric displacement field. Around the van Hove singularities, multiple symmetry-breaking metallic phases are discovered\cite{Zhou2021Half} with reduced spin and valley degrees of freedom, as identified by the doubling or quadrupling of quantum oscillation frequencies. At low temperature, superconducting (SC) phases are further observed\cite{Zhou2021Superconductivity} tightly adjacent to the spin-valley symmetry breaking transitions. In particular, the most prominent SC phase (denoted as SC1 in \refcite{Zhou2021Superconductivity}) appears to be a spin-singlet SC closely proximate to a spin-unpolarized symmetry-breaking metallic phase. Phonon-mediated pairing mechanism\cite{Chou2021Acoustic-phonon-mediated} was soon proposed as a possible scenario. However, it remains unclear why the SC phase follows the boundary of the symmetry-breaking phase so closely and why the enhanced resistivity from electron-phonon scattering is not observed above $T_c$ \cite{Zhou2021Superconductivity}.

\begin{figure}[tbp]
\begin{center}
\includegraphics[width=\columnwidth]{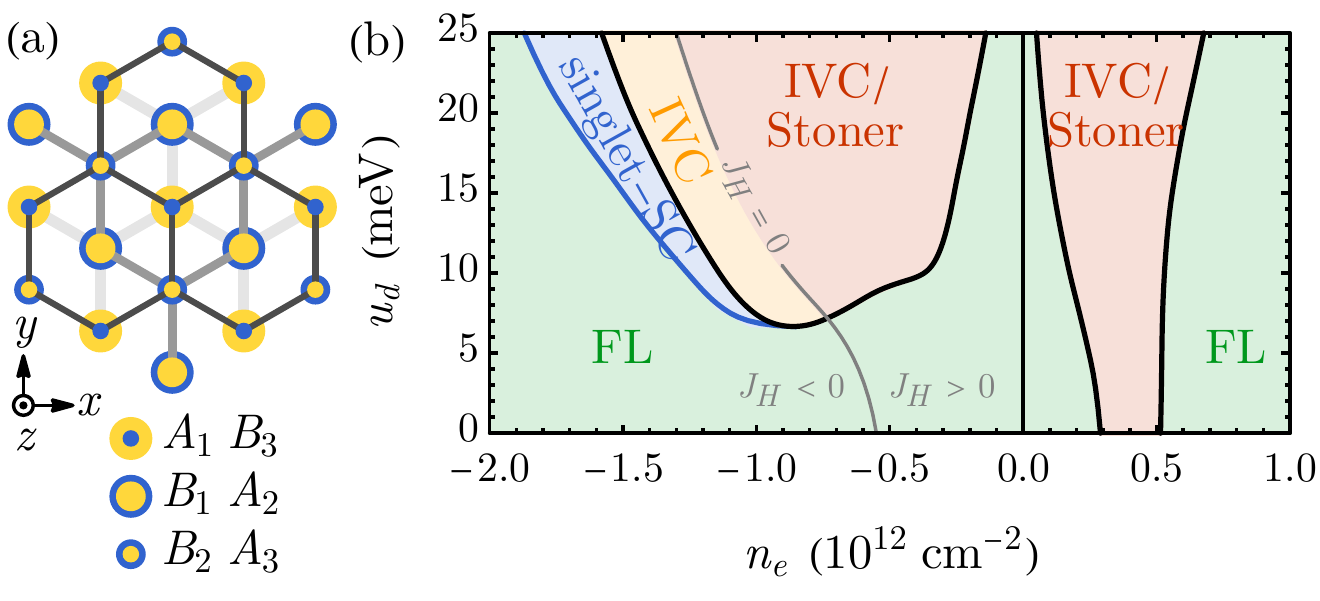}
\caption{(a) Lattice structure (top view) of rhombohedral trilayer graphene. $A_l$ and $B_l$ label the sublattices of the $l$th graphene layer. (b) Schematic phase diagram as a function of density and displacement field, obtained here using different techniques. The IVC phase boundary is set by an RPA instability. Additional spin-valley phase boundaries arising from Stoner instability are not captured.  Singlet superconductivity emerges in proximity to the IVC order with antiferromagnetic Hunds coupling $J_H$. The assume sign change of $J_H$ is schematically shown.}
\label{fig:phase}
\end{center}
\end{figure}

In this work, we investigate an alternative scenario where the SC originates from the Kohn-Luttinger mechanism \cite{Kohn1965New-Mechanism} and its adjacent symmetry-breaking phase is an
inter-valley coherence (IVC) \cite{Po2018Origin,Lee2019Theory,Bultinck2020Ground} phase. Both SC and IVC orders benefit from the inter-valley nesting presented near the van Hove singularity, which naturally explains their tight proximity. Our theory predicts an inter-valley spin-singlet $s$-wave pairing for the SC1 phase with a critical temperature strongly enhanced by the van Hove singularity. \figref{fig:phase}(b) summarizes the schematic phase diagram based on our analysis.

{\it Band Structure Modeling.}~--- The RTG consists of three layers of honeycomb lattices in ABC-stacking as shown in \figref{fig:phase}(a). Its electronic states near charge neutrality mainly reside on $A_1$ and $B_3$ sites, whose low-energy band structure can be described by an effective two-band model\cite{Zhang2010Band,Jung2013Gapped,Ho2016Evolution}
\eq{\label{eq:H0}H_0=\sum_{\vect{k},\xi}c_{\vect{k},\xi}^{\dagger}h_{\xi}(\vect{k})c_{\vect{k},\xi},}
where $c_{\vect{k},\xi}=(c_{\vect{k},\xi,A_1},c_{\vect{k},\xi,B_3})^\intercal$ denotes the electron annihilation operator. $\xi=+$($-$) labels $K$ ($K'$) valley, and $\vect{k}$ labels the momentum deviation from the corresponding valley center. The band Hamiltonian takes the form of
$h_\xi(\vect{k})=(\epsilon_\vect{k}^\text{s}-\mu)\sigma^0+(\xi\alpha_\vect{k}^\text{ch}+\epsilon_\vect{k}^\text{tr})\sigma^1+\beta_\vect{k}^\text{ch}\sigma^2+\epsilon_\vect{k}^\text{gap}\sigma^3$,
with $\alpha_\vect{k}^\text{ch}+\ii \beta_\vect{k}^\text{ch}=\frac{v_0^3}{\gamma_1^2}(k_x+\ii k_y)^3$, $\epsilon_\vect{k}^\text{s}=(\delta+\frac{u_a}{3})-(\frac{2v_0v_4}{\gamma_1}+u_a\frac{v_0^2}{\gamma_1^2})\vect{k}^2$, $\epsilon_\vect{k}^\text{tr}=\frac{\gamma_2}{2}-\frac{2v_0v_3}{\gamma_1}\vect{k}^2$, and $\epsilon_\vect{k}^\text{gap}=u_d(1-\frac{v_0^2}{\gamma_1^2}\vect{k}^2)$. We adopt the parameters proposed in \refcite{Zhou2021Half}, namely $\gamma_0=3.1\text{eV}$, $\gamma_1=380\text{meV}$, $\gamma_2=-15\text{meV}$, $\gamma_3=-290\text{meV}$, $\gamma_4=-141\text{meV}$, $\delta=-10.5\text{meV}$, $u_a=-6.9\text{meV}$, and $v_i=\sqrt{3}a\gamma_i/2$ (for $i=0,3,4$ with $a=0.246\text{nm}$ being the lattice constant). Their physical meanings are well documented in \refcite{Zhang2010Band}. In particular, the parameters $u_d$ and $\mu$ are experimentally tunable by a dual-gate device\cite{Zhou2021Half,Zhou2021Superconductivity}, where $u_d=(u_1-u_3)/2$ is the potential difference between the outer layers (which is approximately proportional to the applied displacement field) and $\mu=-(u_1+u_2+u_3)/3$ is the (overall) chemical potential (assuming $u_l$ to be the electronic potential in the $l$th layer).

\begin{figure}[t]
\begin{center}
\includegraphics[width=0.97\columnwidth]{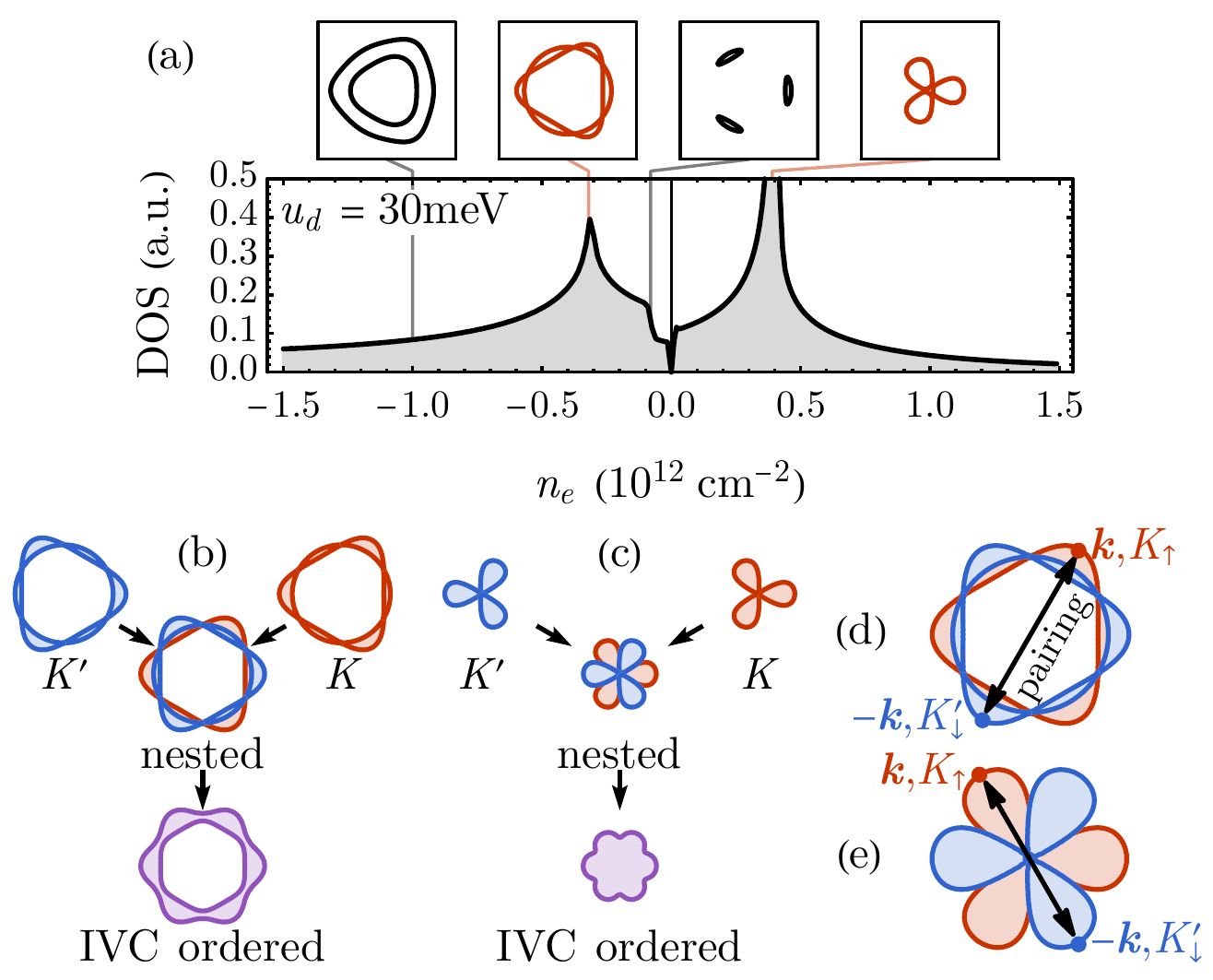}
\caption{(a) Density of state (DOS) v.s. the electron density $n_e$, under a typical displacement field $u_d=30$meV. Insets show the $K$ valley Fermi surfaces at the corresponding doping level (those at van Hove singularities are highlighted in red). (b-c) Ideal scenarios of inter-valley nesting at the van Hove singularity on the (b) hole-doped side and the (c) electron-doped side. The Fermi sea area exactly doubles when the IVC order develops under perfect nesting. (d-e) Perfect inter-valley pairing protected by the time-reversal symmetry on the (d) hole-doped side and the (e) electron-doped side. }
\label{fig:nesting}
\end{center}
\end{figure}

A prominent feature of the RTG band structure is the presence of van Hove singularities in both the conduction and valence bands, where the DOS diverges due to Lifshitz transitions of the Fermi surface topology, as demonstrated in \figref{fig:nesting}(a). On the hole-doped side (in the valence band), the van Hove singularity features an interwoven circular and triangular Fermi surface, whose intersection points are where the larger annular Fermi surface pinches off into smaller Fermi pockets. On the electron-doped side (in the conduction band), the van Hove singularity features a three-leaf clover-shaped Fermi surface when the three smaller pockets are about to merge. These particular shapes of Fermi surfaces at van Hove singularities are well-nested between $K$ and $K'$ valleys, as illustrated in \figref{fig:nesting}(b,c). Under repulsive interactions, the inter-valley nesting leads to a strong instability towards the inter-valley coherence (IVC) order, described by the order parameter $c_{\vect{k},K}^{\dagger}c_{\vect{k},K'}$, which is energetically favored as it gaps out the nested part of the Fermi surface and reduces the DOS at the Fermi level. Furthermore, the time-reversal symmetry ensures that band dispersions around the two valleys are related by $\epsilon_{K'}(\vect{k})=\epsilon_{K}(-\vect{k})$, which admits perfect inter-valley pairing of electrons, as illustrated in \figref{fig:nesting}(d,e). If the interaction is renormalized to attractive, the system will likely develop inter-valley superconductivity in the $s$-wave channel which can fully gap the Fermi surface.

{\it Interaction and Leading Instability.}~--- To analyze these leading instabilities in the RTG system, we introduce the following local interaction of electrons
\eq{\label{eq:Hint}H_\text{int}=\int\dd^2\vect{x}\;\frac{U}{2}(n_{K}^2+n_{K'}^2)+Vn_{K}n_{K'} +H_{\rm Hunds},}
where $n_{\xi}=c_{\xi}^\dagger c_{\xi}$ (for $\xi=K,K'$) denotes the charge operator of the $\xi$ valley. The interaction Hamiltonian $H_\text{int}$ in \eqnref{eq:Hint} takes the most general form that respects the time-reversal, the charge $\U(1)_\mathrm{c}$, valley $\U(1)_\mathrm{v}$ (approximate), and spin $\SU(2)_\mathrm{s}$ symmetries of the band model $H_0$, where $U$ describes the intra-valley density interaction, $V$ describes the inter-valley density interaction, and $H_{\rm Hunds}$ describes the Hund's coupling. At the lattice scale (at the bare level), it is expected that $U,V>0$ are repulsive and $H_{\rm Hunds}$ is small compared to $U,V$ terms. Dropping the Hunds coupling, the symmetry gets enlarged to $\U(1)_\mathrm{c}\times\U(1)_\mathrm{v}\times\SO(4)$, where $\SO(4)\simeq\SU(2)_\mathrm{s}^{K}\times\SU(2)_\mathrm{s}^{K'}$ consists of independent spin rotation symmetries in each valley\cite{You2019Superconductivity}. Further taking $U=V$ will enlarge the symmetry of the interaction to $\U(1)_\mathrm{c}\times\SU(4)$, where valley and spin degrees of freedoms are degenerate\cite{Xu2018Topological}. Of course, the kinetic terms break this SU(4) down to $\U(1)_\mathrm{c}\times\U(1)_\mathrm{v}\times\SO(4)$. 

\begin{table}[htp]
\caption{Complete list of local fermion bilinear operators $\Phi$, classified by symmetry representations of $\U(1)_\mathrm{c}\times\U(1)_\mathrm{v}\times\SO(4)$ following \refcite{You2019Superconductivity}. $g_0^\Phi$ is the bare coupling in each channel (calculated at $H_\text{Hunds}=0$ with $\SO(4)$ symmetry).}
\begin{center}
\begin{tabular}{ccccc}
operator $\Phi$ & $\U(1)_\mathrm{c}$ & $\U(1)_\mathrm{v}$ & $\SO(4)$ & $g_0^\Phi$\\
\hline
$n_c=n_{K}+n_{K'}$ & 0 & 0 & $\mathbf{1}$ & $\frac{1}{4}(U+V)$\\
$n_v=n_{K}-n_{K'}$ & 0 & 0 & $\mathbf{1}'$ & $\frac{1}{4}(U-V)$ \\
$\vect{S}_{\xi}=c_{\xi}^\dagger\vect{\sigma}c_{\xi}$ & 0 & 0 & $\mathbf{6}$ & $-\frac{1}{12}U$\\
$I^\mu=c_{K}^\dagger \sigma^\mu c_{K'}$ & 0 & 2 & $\mathbf{4}\oplus\mathbf{4}'$ & $-\frac{1}{8}V$\\
$\Delta^\mu=c_{K}^\intercal\ii\sigma^2 \sigma^\mu c_{K'}$ & 2 & 0 & $\mathbf{4}\oplus\mathbf{4}'$ & $\frac{1}{8}V$\\
$\Delta_{\xi}=c_{\xi}^\intercal \ii\sigma^2 c_{\xi}$ & 2 & 2 & $2(\mathbf{1}\oplus\mathbf{1}')$ & $\frac{1}{8}U$\\
\end{tabular}
\end{center}
\label{tab:orders}
\end{table}%

\begin{figure}[t]
\begin{center}
\includegraphics[width=0.9\columnwidth]{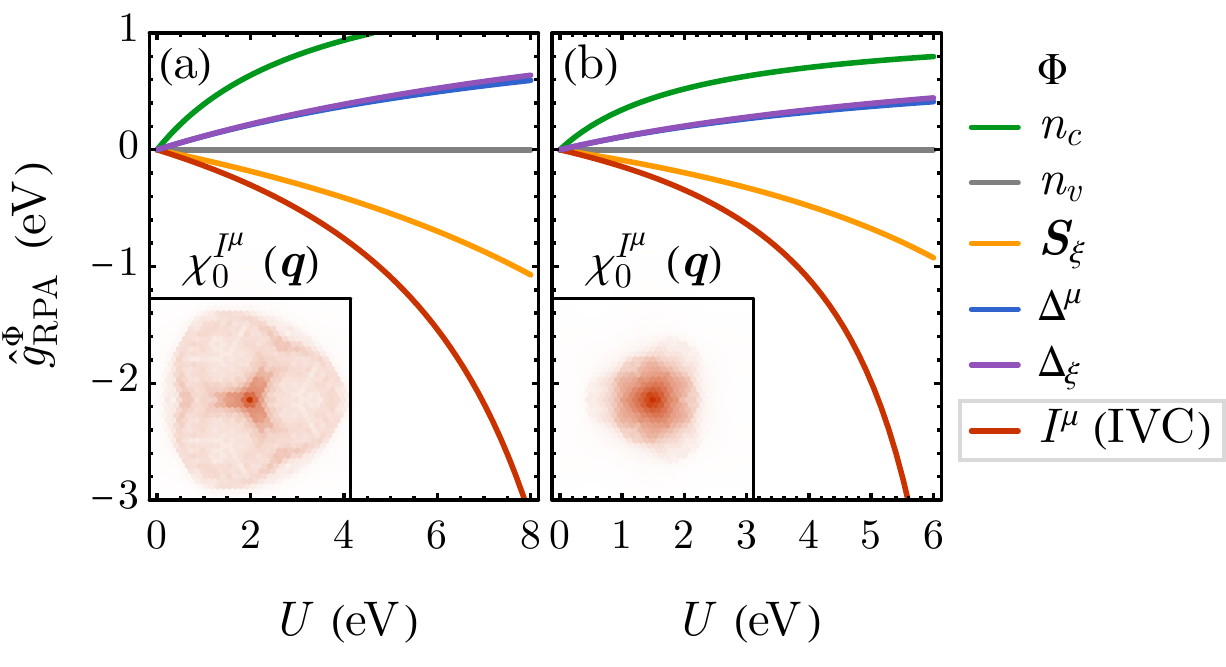}
\caption{The peak value of RPA corrected coupling $\hat{g}_\text{RPA}^{\Phi}$ in different ordering channels v.s. the bare interaction strength $U=V$ under the displacement field $u_d=30$meV at the van Hove singularity on (a) the hole-doped side and (b) the electron-doped side. The IVC channel is always the leading instability. The insets shows the wavevector dependence of the bare susceptibility $\chi_0^{I^\mu}(\vect{q})$ in the IVC channel.}
\label{fig:RPA}
\end{center}
\end{figure}

For simplicity, we take the interaction to be repulsive and $\SU(4)$ symmetric by setting $U=V>0$ and $H_{\rm Hunds}=0$, and apply the random phase approximation (RPA)\cite{Kuroki2008Unconventional,Graser2009Near,Maier2011} approach to analyze the Fermi surface instabilities towards all possible local fermion bilinear orders $\Phi$, as enumerated in \tabref{tab:orders} (strictly speaking, the total charge density $n_c$ can not order spontaneously, as it is controlled by the chemical potential). For each fermion bilinear operator $\Phi_{\vect{q}}=\frac{1}{2}\sum_{\vect{k}}\chi_{-\vect{k}+\vect{q}}^\intercal \Phi\chi_{\vect{k}}$ generally expressed in the Majorana basis $\chi_\vect{k}$, we evaluate its bare static (zero frequency) susceptibility $\chi_{0}^{\Phi}(\vect{q})\equiv\langle \Phi_{\vect{q}}^\dagger \Phi_{\vect{q}}\rangle_0$ on the thermal equilibrium state of the band Hamiltonian $H_0$ at a low temperature $T=0.29$K ($25\upmu$eV). Then we rewrite the interaction $H_\text{int}=g_0^{\Phi}\sum_{\vect{q}}\Phi_{\vect{q}}^\dagger \Phi_{\vect{q}}+\cdots$ in the same channel to extract the bare coupling $g_0^\Phi$ (see the last column of \tabref{tab:orders}). The RPA corrected coupling is then given by $g_\text{RPA}^{\Phi}(\vect{q})=g_0^{\Phi} (1+g_0^{\Phi}\chi_0^{\Phi}(\vect{q}))^{-1}$. A strong negative (attractive) coupling $g_\text{RPA}^{\Phi}$ indicates a strong ordering tendency of the corresponding order parameter $\Phi$. The peak value $\hat{g}_\text{RPA}^{\Phi}\equiv\mathop{\mathrm{max}}_\vect{q} g_\text{RPA}^{\Phi}(\vect{q})$ is taken and plotted in \figref{fig:RPA} as a function of the bare interaction $U$ in various channels, at the van Hove singularity on the (a) hole-doped and (b) electron-doped sides. In either cases, the most attractive coupling appears in the IVC channel, which is associated with the operator $I^\mu_\vect{q}=\sum_{\vect{k}}c_{\vect{k}+\vect{q},K}^\dagger \sigma^\mu c_{\vect{k},K'}$. Comparing \figref{fig:RPA}(a) and (b), the electron-doped side has a stronger IVC instability, which is consistent with the stronger van Hove singularity on the electron-doped side, as shown in \figref{fig:nesting}(b). We would like to comment that the RPA analysis only captures the nesting/pairing instability. The spin or valley polarization ($\vect{S}_\xi$ or $n_v$ ordering) under Stoner instability is outside the scope of RPA calculation. How the IVC order may compete with the spin-valley polarization will be left for future study.

{\it Inter-Valley Coherence.}~--- The momentum distributions of the bare IVC susceptibility $\chi_0^{I^\mu}(\vect{q})$ are shown in the insets of \figref{fig:RPA}, which peaks at $\vect{q}=\vect{0}$ under optimal nesting near van Hove singularities. Therefore, when the bare interaction is strong enough, $I^\mu_{\vect{q}}$ will condense at zero momentum (modulo the inter-valley momentum), leading to a uniform IVC order at a large scale. Nevertheless, the IVC order does carry the inter-valley momentum difference, which should appear as a charge/spin density wave (CDW/SDW) with Kekul\'e pattern on the lattice scale\cite{Halperin1968Possible,Aleiner2007Spontaneous}. In the absence of the $\SO(4)$ symmetry-breaking interaction $H_{\rm Hunds}$, the four components of the IVC order $I^\mu$ ($\mu=0,1,2,3$) are degenerate as an $\SO(4)$ vector, where $I^0$ is the spin-singlet (CDW-like) IVC and $\vect{I}=(I^1,I^2,I^3)$ is the spin-triplet (SDW-like) IVC. This degeneracy can be split by the Hund's interaction either as the spin coupling $H_\text{Hunds}=-J_H\vect{S}_{K}\cdot\vect{S}_{K'}$ or as the IVC coupling $H_\text{Hunds}=-\tilde{J}_H\vect{I}^\dagger\cdot\vect{I}$ (note that the two types of couplings are related by $\vect{S}_{K}\cdot\vect{S}_{K'}=2\vect{I}^\dagger\cdot\vect{I}+3n_{K}n_{K'}$ in the local interaction limit when the couplings are momentum independent). For example, an antiferromagnetic coupling ($J_H<0$ or $\tilde{J}_H<0$) will favor the spin-singlet IVC. The spin-singlet IVC order could be directly spotted in the scanning tunneling microscopy (STM) image\cite{Li2019Scanning}. 

\begin{figure}[t]
\begin{center}
\includegraphics[width=0.8\columnwidth]{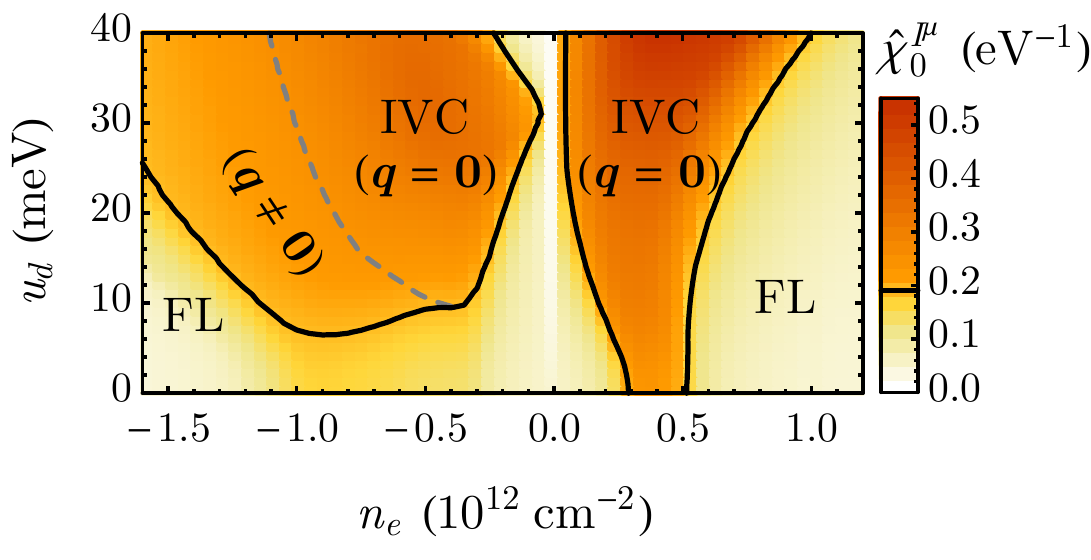}
\caption{The maximal bare static IVC susceptibility $\hat{\chi}_0^{I^\mu}$ as a function of the electron density $n_e$ and the displacement field $u_d$, calculated at temperature $T=1.8$K ($155\upmu$eV). The mean-field phase boundary between the IVC phase and the Fermi liquid (FL) phase is determined by assuming the inter-valley density interaction to be $V=42$eV (at lattice scale). Inside the hole-doped IVC phase, the dashed curve separates the IVC order of zero momentum from that of finite momentum.}
\label{fig:IVCphase}
\end{center}
\end{figure}

Within the RPA approach, the ordering transition happens when the effective coupling $g_\text{RPA}^\Phi(\vect{q})$ diverges at some momentum $\vect{q}$ where $1+g_0^\Phi\chi_0^\Phi(\vect{q})=0$. This leads to the mean-field equation for the IVC order
\eq{\label{eq:mean-field}\hat{\chi}_0^{I^\mu}\equiv\max_\vect{q}\chi_0^{I^\mu}(\vect{q})=-\frac{1}{g_0^{I^\mu}}=\frac{8}{V}.}
As shown in \figref{fig:IVCphase}, the maximal bare static susceptibility $\hat{\chi}_0^{I^\mu}$ varies with both the displacement field $u_d$ and the chemical potential $\mu$ (where $\mu$ can be translated to the electron density $n_e$ assuming a rigid band). By solving the mean-field equation \eqnref{eq:mean-field}, the IVC phase boundary can be traced out in the $n_e$-$u_d$ plane, see \figref{fig:IVCphase}. The phase diagram looks most similar to the experiment\cite{Zhou2021Half} by taking $V\approx 42$eV \footnote{Note the analogous Stoner parameter used in Ref \cite{Zhou2021Half} was in the same ball park $\sim 30$eV}. Near the van Hove singularities ($n_e\sim\pm0.5\times10^{12}\text{cm}^{-2}$), uniform ($\vect{q}=\vect{0}$) IVC order is favored due to good nesting. Doping away from the van Hove singularity (on the hole-doped side), the IVC order can shift to finite momentum $\vect{q}$. The finite-$\vect{q}$ IVC order will either break the three-fold lattice rotation symmetry (if the IVC order parameter condenses at a single $\vect{q}$ point \cite{kwan2021kekule}), or break the (larger-scale) translation symmetry (if it condenses to symmetry-related $\vect{q}$ points coherently \cite{You2019Superconductivity}).

\begin{figure}[t]
\begin{center}
\includegraphics[width=0.93\columnwidth]{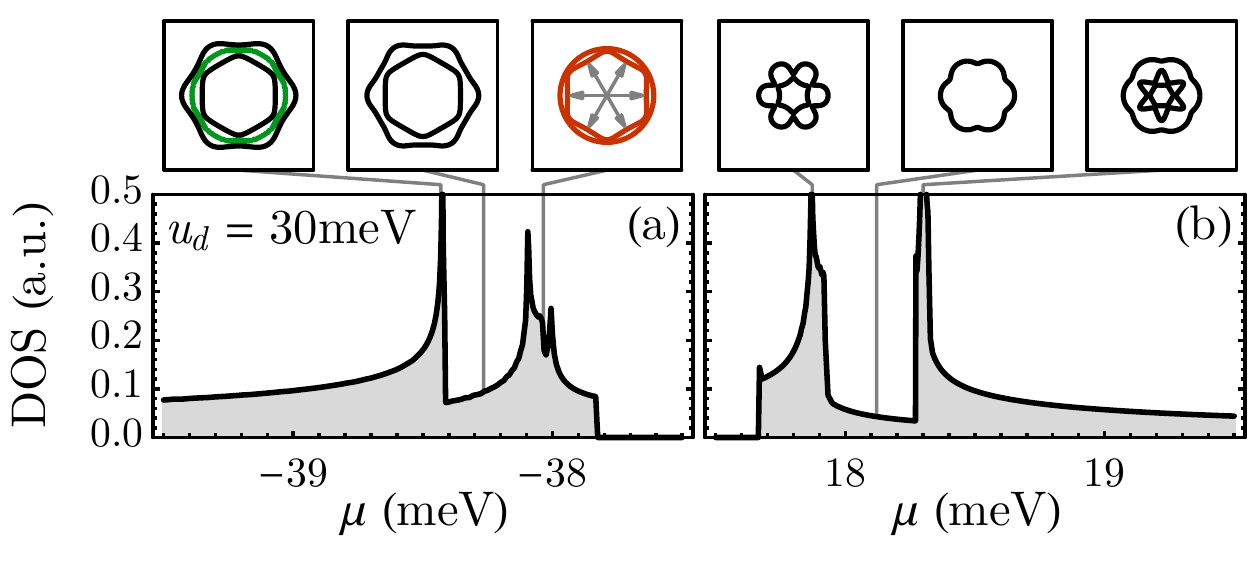}
\caption{Density of state (DOS) and Fermi surface topology under a uniform ($\vect{q}=\vect{0}$) IVC order that opens $\sim0.4\text{meV}$ partial gap near the original van Hove singularities on (a) the hole-doped side and (b) the electron-doped side.}
\label{fig:IVC}
\end{center}
\end{figure}

As shown in \figref{fig:IVC}, the IVC ordering splits the van Hove singularities in the original band structure, and significantly reduces the Fermi surface DOS. However, it does not fully gap out the Fermi-level degrees of freedom. Instead, it merges the two sets of Fermi surfaces from separate valleys to create a single set of Fermi surfaces shared between valleys, as illustrated in \figref{fig:nesting}(c,d). Thus the IVC order effectively locks the valley degrees of freedom together and doubles the Fermi sea area, which provides a possible explanation for the doubling of quantum oscillation frequency (from $f_\nu=1/4$ to $1/2$) observed in experiment\cite{Zhou2021Half}. As the system remains metallic, it is further subject to other Fermi surface instability. For example, the hexagonal inner Fermi surface (highlighted in red in \figref{fig:IVC}) near the hole-under-doped van Hove singularity is well-nested between parallel sides (as indicated by arrows), which may lead to a secondary SDW order on top of the IVC order, such that the quantum oscillation frequency could be further doubled (to $f_\nu=1$). While near the hole-over-doped van Hove singularity, the circular Fermi surface (highlighted in green in \figref{fig:IVC}) exhibits a strong pairing instability, which could potentially give rise to superconductivity.

{\it Superconductivity.}~--- It is recently discovered in experiment \cite{Zhou2021Superconductivity} that superconducting (SC) phases emerge at low temperature within the Fermi liquid (FL) phase, and are closely adjacent to the boundary of the spin-valley symmetry breaking phase (which we identified as the IVC phase). However, signatures of enhanced scattering from the underlying pairing-glue modes is not observed above the SC transition temperature \cite{Zhou2021Superconductivity}, which points to the Kohn-Luttinger mechanism\cite{Kohn1965New-Mechanism,Raghu2010Superconductivity,Maiti2013Su,Kagan2015Anomalous,Chubukov2017Superconductivity}
for superconductivity based on the intrinsic instability of the FL. To explore this possibility, we perform the patch renormalization group (RG) analysis \cite{Furukawa1998Truncation,Raghu2011Superconductivity,Isobe2018Unconventional,Nandkishore2012Chiral,Park2021Electronic, Betouras} that tracks the RG flow of interactions at one-loop level. We will focus on the van Hove singularity in a single band, described by $H_0=\sum_{\vect{k},\xi}c_{\vect{k},\xi}^\dagger\epsilon_{\xi}(\vect{k})c_{\vect{k},\xi}$, where the band dispersion $\epsilon_{\xi}(\vect{k})$ can be obtained by diagonalizing the two-band model in \eqnref{eq:H0} and pick one band of interest. It will be convenient to decompose the dispersion $\epsilon_\xi(\vect{k})=\xi \pi_{\vect{k}}+\mu_{\vect{k}}$ in terms of the time-reversal even $\mu_{\vect{k}}=\mu_{-\vect{k}}$ and odd $\pi_{\vect{k}}=-\pi_{-\vect{k}}$ components, and redefine $\epsilon_\vect{k}\equiv\pi_\vect{k}+\mu_\vect{k}$. 

We consider the Hund's interaction in the form of inter-valley spin coupling $H_\text{Hunds} = -J_H\vect{S}_K\cdot \vect{S}_{K'}$ projected onto states near the Fermi surface. Note, $J_H<0$ corresponds to antiferromagnetic Hund's interaction. By treating the two valleys as two patches in the patch RG framework, we can employ the general approach developed in \refcite{Furukawa1998Truncation} to renormalize the interactions in \eqnref{eq:Hint}, and obtain the following RG equations (see \appref{app:RGEq} for derivation)
\eqs{\label{eq:RG}
\tfrac{\dd}{\dd y}U&=-\tfrac{1}{2}d_{\vect{0}}^\text{pp}U^2,\\
\tfrac{\dd}{\dd y}V&=-\tfrac{1}{2}(1-d_{\vect{Q}}^\text{ph})(V^2+3J_H^2),\\
\tfrac{\dd}{\dd y}J_H&=-J_H((1+d_{\vect{Q}}^\text{ph})J_H+(1-d_{\vect{Q}}^\text{ph})V),\\}
where the RG ``time'' $y=\Pi_{\vect{Q}}^\text{pp}(E)$ and the nesting parameters $d_{\vect{0}}^\text{pp}(y)=\partial_y\Pi_{\vect{0}}^\text{pp}(E)$, $d_{\vect{Q}}^\text{ph}(y)=\partial_y\Pi_{\vect{Q}}^\text{ph}(E)$ are related to the particle-particle (pp) or particle-hole (ph) ladder diagram $\Pi$ down to the energy scale $E$ at the momentum transfer of either $\vect{0}$ (intra-valley) or $\vect{Q}$ (inter-valley). More explicitly, in the zero-temperature limit, $\Pi_{\vect{Q}}^\text{pp}(E)=\sum_{|\epsilon_\vect{k}|>E}\frac{1}{|\epsilon_\vect{k}|}$, $\Pi_{\vect{0}}^\text{pp}(E)=\sum_{|\mu_\vect{k}|>E}\frac{\Theta(|\mu_\vect{k}|-|\pi_\vect{k}|)}{|\mu_\vect{k}|}$, $\Pi_{\vect{Q}}^\text{ph}(E)=\sum_{|\pi_\vect{k}|>E}\frac{\Theta(|\pi_\vect{k}|-|\mu_\vect{k}|)}{|\pi_\vect{k}|}$. At the van Hove singularity (assuming the density of state diverges as $2\nu_0\ln(\Lambda/\epsilon)$ with $\Lambda$ being the UV cuttoff), the RG time will increase double logarithmically $y=\nu_0\ln^2(\Lambda/E)$ with the decreasing energy scale $E$, while the nesting parameters vary within the range $0\leq d_{\vect{0}}^\text{pp},d_{\vect{Q}}^\text{ph}\leq 1$ and typically decay with $y$. Similar RG equations can be derived, if the Hund's interaction $H_\text{Hunds}=-\tilde{J}_H\vect{I}^\dagger\cdot\vect{I}$ takes the form of IVC coupling (see \appref{app:Hunds}).

\begin{figure}[t]
\begin{center}
\includegraphics[width=0.95\columnwidth]{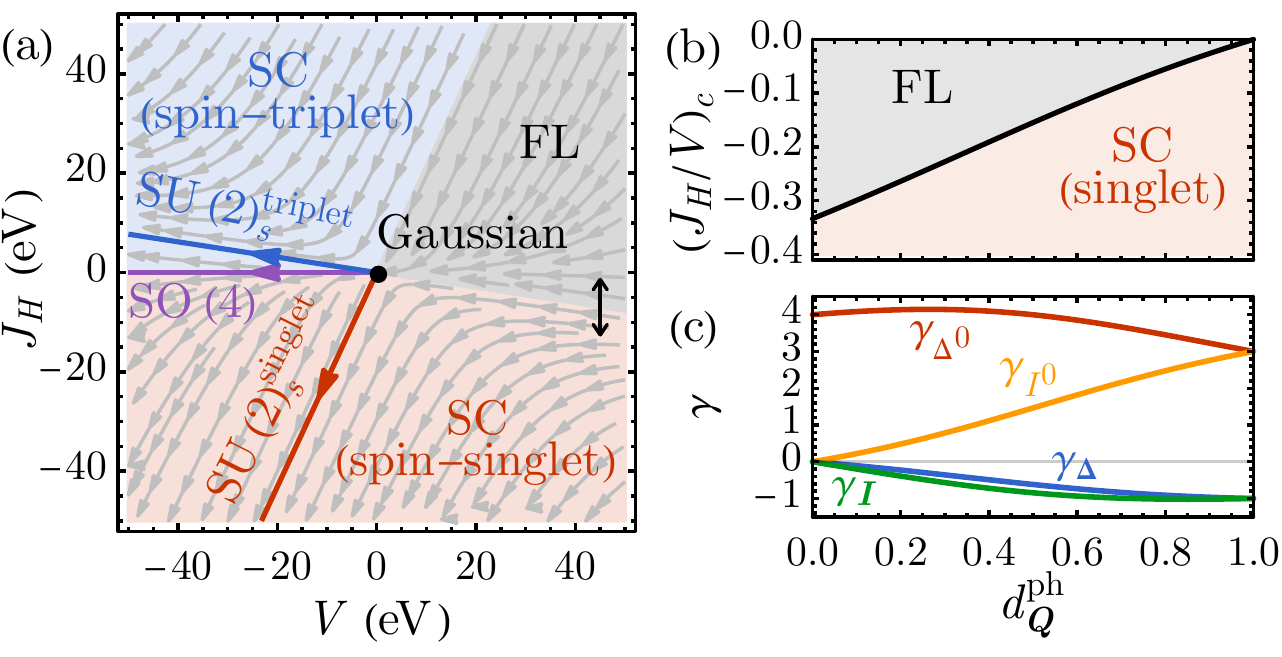}
\caption{(a) RG flow diagram of $(V,J_H)$ with a typical choice of the nesting parameter $d_\vect{Q}^\text{ph}=0.5$. Fixed rays/points of RG flows are classified by symmetries. They separately control different superconducting (SC) and Fermi liquid (FL) phases. (b) The critical $|J_H/V|$ ratio for the spin-singlet SC and FL transition as a function of $d_\vect{Q}^\text{ph}$. (c) Critical exponents for singlet ($\gamma_{\Delta^0}$) and triplet-paring ($\gamma_{\vect{\Delta}}$) susceptibility at the SC transition, as well as exponents for IVC orders ($\gamma_{I^0}$, $\gamma_{\vect{I}}$).}
\label{fig:RG}
\end{center}
\end{figure}

For repulsive interaction, the intra-valley density interaction $U$ is always marginally irrelevant and decouples from the flow of $V$ and $J_H$, so we only need to focus on the RG flow in the $V$-$J_H$ plane. A typical flow diagram is shown in \figref{fig:RG}(a). We first look at the $J_H=0$ axis, where the interaction is $\SO(4)$ symmetric. Based on the RG equation \eqnref{eq:RG}, the inter-valley density interaction $V$ always decrease under RG, which will either flow to the Gaussian (free) fixed point at $V=0$ (if $V>0$ initially), or flow away towards $V\to-\infty$ (if $V<0$ initially) along the $\SO(4)$ fixed ray in \figref{fig:RG}(a). A strong and negative $V$ will drive an instability towards the inter-valley SC pairing $\Delta^\mu=c_{K}^\intercal\ii\sigma^2 \sigma^\mu c_{K'}$ ($\mu=0,1,2,3$) presumably in the $s$-wave channel (as $\Delta^\mu$ gains energy the most from a negative $V$ interaction due to the $\frac{V}{8}|\Delta^\mu|^2$ effective potential listed in \tabref{tab:orders}). Under the $\SO(4)$ symmetry, the spin-singlet $\Delta^0$ and spin-triplet $\vect{\Delta}=(\Delta^1,\Delta^2,\Delta^3)$ pairings are degenerate. However, the degeneracy can be split by the Hund's interaction $H_\text{Hunds}$, which could either be ferromagnetic ($J_H>0$) or antiferromagnetic ($J_H<0$) depending on microscopic details\cite{Chakravarty1991,Kivelson2018}. In any case, the $\SO(4)$ fixed ray is unstable under the perturbation of $J_H$, and flows to either the $\SU(2)_\text{s}^\text{triplet}$ or $\SU(2)_\text{s}^\text{singlet}$ fixed rays in \figref{fig:RG}(a), which separately control the spin-triplet or spin-singlet inter-valley SC phases. 

Although the $\SO(4)$-breaking interaction $J_H$ might be small compare to $U$ and $V$, it plays an important role as a dangerously irrelevant perturbation to help an initially repulsive ($V>0$) interaction to evade the attraction of the Gaussian fixed point and to change the sign to attractive ($V<0$) that enables the $s$-wave SC order. However, the bare $(J_H/V)$ ratio (on the $V>0$ side) must exceed a certain critical value (in terms of magnitude) to achieve this effect. Let us focus on the transition between spin-singlet SC phase and the FL phase, as marked by the two-away arrow in \figref{fig:RG}(a), the corresponding critical ratio is given by $(J_H/V)_c=d_V/d_J^-$ where $d_V=1-d_\vect{Q}^\text{ph}$ and $d_J^{\pm}=\pm2(1-d_\vect{Q}^\text{ph}+(d_\vect{Q}^{\text{ph}})^2)^{1/2}-1-d_\vect{Q}^\text{ph}$, which can be reasonably small if the inter-valley nesting is almost perfect (i.e.~$d_\vect{Q}^\text{ph}\to 1$), see \figref{fig:RG}(b). Notice that a good nesting is also the condition for IVC instability on the mean-field level, which explains why the SC phase (e.g.~SC1 in \refcite{Zhou2021Superconductivity}) comes so close to the boundary of the IVC phase in the experiment, as both phases want to take advantage from the inter-valley nesting. However, on the spin-triplet SC side, the critical ratio is given by $(J_H/V)_c=d_V/d_J^+$, which is bounded by $(J_H/V)_c \geq 1$ for all range of $d_\vect{Q}^\text{ph}$. Therefore, if we assume $J_H$ to be small compare to $V$, the spin-triplet SC will be unlikely to develop inside the FL phase (unless adjacent to other FM spin-ordered phases).

Focusing on the spin-singlet SC phase for $(J_H/V)<(J_H/V)_c$, the RG flow will diverge at some critical scale $y_c=1/V_\text{eff}$ along the $\SU(2)_\text{s}^\text{singlet}$ fixed ray, where $V_\text{eff}=\frac{1}{2}(d_J^- J_H-d_V V)>0$. This sets a critical temperature
\eq{T_c\sim\Lambda \exp(-1/\sqrt{\nu_0V_\text{eff}}).}
The $\sqrt{V_\text{eff}}$ dependence was highlighted in several existing literature\cite{Son1999Superconductivity,Moon2010Quantum-critical,Nandkishore2012Chiral}. It results in a $T_c$ that is strongly enhanced compared to the standard Bardeen-Cooper-Schrieffer (BCS) result $T_c\sim \exp(-1/\nu_0V_\text{eff})$, which could put the proposed Kohn-Luttinger mechanism to work in the experimentally accessible temperature range. 

To determine the nature of the ordered phase below $T_c$, we examine how the susceptibility $\chi^\Phi\equiv\langle\Phi^\dagger \Phi\rangle\sim(y_c-y)^{-\gamma_\Phi}$ diverges near the critical scale $y_c$ by computing the critical exponent $\gamma_\Phi$ for various order parameters $\Phi$ (as listed in \tabref{tab:orders}). Following \refcite{Furukawa1998Truncation,Nandkishore2012Chiral} (see \appref{app:scaling} for details), along the $\SU(2)_\text{s}^\text{singlet}$ fixed ray, we find $\gamma_{n_c}=\gamma_{n_v}=\gamma_{\vect{S}_\xi}=\gamma_{\Delta_\xi}=0$ with the remaining non-trivial exponents $\gamma_{I^\mu},\gamma_{\Delta^\mu}$ plotted in \figref{fig:RG}(c). The spin-singlet inter-valley pairing $\Delta^0$ has indeed the largest exponent $\gamma_{\Delta^0}$ (hence the fastest divergence) as the leading instability, confirming the proposed SC phase below $T_c$. We also note that the CDW-type IVC order $I^0$ is secondary and becomes degenerate with $\Delta^0$ at perfect nesting. The RG analysis assumes infinitesimal interactions. When the bare interaction is small but not close to the fixed ray, sub-leading instability (such as the IVC order) may first occur at the mean-field level before the RG can reverse the sign of $V$ and trigger the SC instability, which is compatible with the RPA result.

Another remarkable phenomenon is that the small $\SO(4)$ symmetry breaking interaction $J_H$ can be amplified by the RG flow. As seen in \figref{fig:RG}(c), both the spin-triplet SC and spin-triplet IVC orders ($\vect{\Delta}$ and $\vect{I}$) are suppressed under RG (with negative exponents), which is consistent with the experimental observation that no spin-triplet SC remains when the spin-singlet SC is suppressed by an external in-plane magnetic field (i.e. Pauli limited superconductivity\cite{Clogston1962Upper,Chandrasekhar1962a-Note}). 

{\it Discussion.}~--- Before closing, let us highlight some issues raised by experiments. In the vicinity of SC1 we have assumed antiferromagnetic spin coupling $J_H<0$,  \cite{Zhou2021Superconductivity}
. The same sign of $J_H$ is observed in other graphene structures, e.g.~monolayers and bilayers \cite{Maher_2013,Maher2013Evidence}. However, the spin-split Fermi surfaces at smaller hole doping points to $J_H>0$, implying that it changes sign with doping, as illustrate in \figref{fig:phase}(b). Clarifying this evolution will require a microscopic calculation of $J_H$, an important future task. Across the $J_H=0$ line, the IVC phase can be split into the spin-singlet ($J_H<0$) and spin-triplet ($J_H>0$) sub-phases, see \figref{fig:phase}  A different mechanism for IVC fluctuation superconductivity was discussed in \cite{You2019Superconductivity}, even when SO(4) symmetry is preserved. There, topological superconductivity was obtained and the details of that mechanism are rather different from the one we explored here. We propose that the spin unpolarized PIP phase could correspond to an IVC phase. However, our RPA approach does not capture the Stoner instability of spin or valley polarized states, which is another competitive valley-spin symmetry breaking order near the van Hove singularity. We will leave the competition between IVC and Stoner ferromagnetism for future investigation.

Next, we note that the observation of a Pauli limiting field in SC1 is significant, particularly if we assume that the in-plane magnetic field couples entirely through the Zeeman effect. As a side remark, although it is natural to ascribe the observed Pauli limiting behavior to the Zeeman coupling, we note that potentially, an in-plane field could also lead to an orbital suppression of superconductivity, due to finite layer separation. 
Future experiments probing in-plane field anisotropy, expected for the orbital but not for the Zeeman coupling, should help isolate the relevant coupling. At present, the absence of any in-plane field effect on the putative triplet superconductor SC2, and the observation of the Pauli limit in SC1, suggests pure Zeeman coupling.) Under this assumption, the Pauli limiting behavior strongly suggests that anisotropy terms like $J_H$ play a significant role in superconductivity. If instead pairing was primarily determined by $\SO(4)$ symmetric interactions, then one would expect both singlet and triplet superconductors to be nearly degenerate. The $\SO(4)$ breaking anisotropies would lead only to small splittings in $T_c$ implying that on destroying the spin-singlet superconductor with Zeeman field, a spin-triplet superconductor, with slightly reduced $T_c$ would be established beyond the Pauli limit. Indeed, such behavior is observed in a moir\'e material of mirror-symmetric twisted trilayer graphene\cite{Cao2021Pauli-limit}, but is not observed in the RTG SC1 phase. This indicates a potentially different origin for superconductivity here compared to moir\'e graphene. Our pairing mechanism emphasizes the role of $J_H$ from the start, in contrast to those that rely predominantly on $\SO(4)$ symmetric interactions, either electronic or those based on acoustic phonons. The Hund's coupling $J_H$ plays an indispensable role to allow the inter-valley density interaction $V$ to change sign under RG, which enables the $s$-wave SC to emerge even under a repulsive bare interaction. A weak coupling $J_H$ also flows strong under RG, which explains the absence of spin-triplet SC phase above the Pauli limit, even if the $\SO(4)$ symmetry is only weakly broken at the bare level.

Finally, it is interesting to compare and contrast magic-angle graphene (bilayer and mirror-symmetric trilayer) and RTG. On one hand, differences include the fact that the transition temperatures in the former are larger than the latter by over an order of magnitude despite comparable electronic densities, and show signatures of strong coupling superconductivity \cite{Cao2018a,Yankowitz2018Tu,Lu2019Superconductors,Park2021Tunable,Hao2021Electric} along with strange metal normal state behavior \cite{Cao2020Strange, Polshyn2019Large, Jaoui2021Quantum-critical}. The width in doping supporting superconductivity is also dramatically different in the two cases. However, we note that the proximity of isospin order and superconductivity is a common feature in both. Comparing these two remarkable superconductors will be a useful future exercise. 

{\it Summary.}~--- We combine the mean-field/RPA approach and the patch RG analysis to investigate the IVC and inter-valley SC orders in RTG systems. The particular Fermi surface topology at the van Hove singularity enables a good inter-valley nesting, which promotes both the IVC and SC ordering under local electronic interactions. The IVC picture nicely reproduces the general shape of the phase diagram and explains the doubling of quantum oscillation frequency. It also leaves room for cascades of secondary ordered phases when the system is further doped away from the van Hove singularity. The diverging density of state near the van Hove singularity also strongly renormalizes the interaction in the FL phase, which, particularly in the presence of an inter-valley antiferromagnetic $J_H<0$ coupling, causes the inter-valley density interaction to become attractive at low-energy, favoring the inter-valley spin-singlet $s$-wave SC with an enhanced critical temperature $T_c$ compared to the standard BCS formula. The RG flow can also amplify the effect of the inter-valley Hunds interaction, leading to a significant splitting of spin-singlet and spin-triplet SC states, even if the $\SO(4)$ breaking anisotropy is weak at the lattice scale.

{\it Notes Added:} While completing this manuscript, related works Refs. \cite{Ghazaryan2021Unconventional,Chatterjee2021Inter-valley,Dong2021Superconductivity,Szabo2021Parent,Cea2021Superconductivity} appeared, which propose alternative pairing mechanisms and sign changing or topological superconductivity. In particular Ref. \cite{Chatterjee2021Inter-valley} identified an IVC ground state through a microscopic calculation.

{\it Acknowledgements:} We thank Fa Wang, Mike Zaletel, Taige Wang, and especially Andrea Young for stimulating discussions. YZY was supported by a startup fund provided by UCSD and a UC Hellmann Fellowship. AV was supported by a Simons Investigator award.  

\bibliography{refs}

\pagebreak
\onecolumngrid
\appendix
\section{RG Analysis}
\subsection{Derivation of RG Equation}\label{app:RGEq}
We start with the local interaction of electrons (here we take a Hund's coupling in the form of spin coupling and carry out the derivation, we will discuss the other type of Hund's coupling in the form of IVC coupling later)
\eq{\label{eq:H_int_supp}H_\text{int}=\int\dd^2\vect{x}\;\frac{U}{2}(n_{K}^2+n_{K'}^2)+Vn_{K}n_{K'}-J_H\vect{S}_{K}\cdot\vect{S}_{K'},}
where $n_{\xi}=c_{\xi}^\dagger c_{\xi}$ and $\vect{S}_{\xi}=c_{\xi}^\dagger\vect{\sigma}c_{\xi}$ (for $\xi=K,K'$) are respectively the charge and spin density operators near the $\xi$ valley, and $c_\xi=(c_{\xi\uparrow},c_{\xi\downarrow})^\intercal$ is the electron annihilation operator. To derive the RG equation more systematically, it will be convenient to rewrite the interaction \eqnref{eq:H_int_supp} in the Majorana fermion basis (by rewriting the electron operator as $c_{\xi\sigma}=(\chi_{\xi\sigma 1}+\ii\chi_{\xi\sigma 2})/\sqrt{2}$),
\eq{H_\text{int}=\int\dd^2\vect{x}\;\sum_{a,b,c,d}V_{abcd}\chi_a\chi_b\chi_c\chi_d,}
where $a,b,c,d=1,\cdots,8$ runs over eight Majorana components (valley$\times$spin$\times$particle-hole), and $V_{abcd}$ is a totally antisymmetric tensor. In particular, its representative components are given by
\eqs{\label{eq:V}&V_{1234}=V_{5678}=-U,\\
&V_{1256}=V_{3478}=-V+J_H,\\
&V_{1278}=V_{3456}=-V-J_H,\\
&V_{1357}=V_{1458}=V_{2367}=V_{2468}=V_{1368}=V_{1476}=V_{2385}=V_{2457}=J_H.}
The advantage of using the Majorana basis is that the interaction vertex $V$ receives one-loop correction via a single unified ladder diagram (which unifies all types of ladder/bubble/vertex-correction diagrams in the complex fermion basis). By perturbative expansion, the corrected interaction vertex at the energy scale $E$ is given by
\eq{V_{abcd}(E)=\raisebox{-15pt}{\includegraphics[width=20pt]{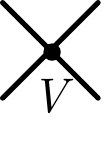}}+\sum_{k\in\scK_E}
\raisebox{-20pt}{\includegraphics[width=60pt]{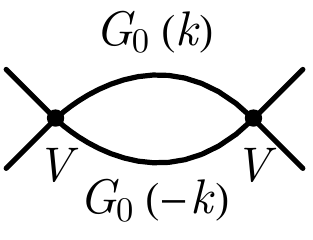}}+\cdots=V_{abcd}-V_{abc'd'}\Pi_{c'd'a'b'}(E)V_{a'b'cd}+\cdots,}
where $\scK_{E}$ denotes the frequency-momentum space beyond the energy scale $E$. Here $\Pi(E)$ denotes the ladder kernel at the energy scale $E$ (by integrating out virtual processes beyond the scale $E$), 
\eq{\label{eq:Pi_def}\Pi_{c'd'a'b'}(E)=\sum_{k\in\scK_E}G_0(k)_{c'b'}G_0(-k)_{d'a'},}
where $G_0(k)$ denotes the bare fermion propagator in the Majorana basis. Upon the reduction of the energy scale $E\to E-\dd E$, the RG equation will be given by
\eq{\label{eq:dV}\dd V_{abcd}=-V_{abc'd'}\dd\Pi_{c'd'a'b'}V_{a'b'cd}.}

To evaluate ladder kernel $\Pi(E)$, we take a single-band model in the momentum space (we assume that the interaction has been projected to the band of interest)
\eq{\label{eq:H0_supp}H_0=\sum_{\vect{k},\xi}c_{\xi,\vect{k}}^\dagger \epsilon_{\xi,\vect{k}} c_{\xi,\vect{k}},}
where $\xi$ labels the $K$ and $K'$ valleys, and $\epsilon_{\xi,\vect{k}}$ be the band dispersion near the $\xi$ valley. By the time-reversal symmetry, the band dispersion $\epsilon_{K,\vect{k}}$ near the $K$ valley must be related to that $\epsilon_{K',\vect{k}}$ near the $K'$ valley by $\epsilon_{K',\vect{k}}=\epsilon_{K,-\vect{k}}$. Let us define $\epsilon_{K,\vect{k}}=\epsilon_{\vect{k}}=\pi_{\vect{k}}+\mu_{\vect{k}}$, in terms of the odd $\pi_{-\vect{k}}=-\pi_{\vect{k}}$ and even $\mu_{-\vect{k}}=\mu_{\vect{k}}$ parity parts. Then the dispersion around the $K'$ valley can be expressed as $\epsilon_{K',\vect{k}}=\epsilon_{K,-\vect{k}}=\pi_{-\vect{k}}+\mu_{-\vect{k}}=-\pi_{\vect{k}}+\mu_{\vect{k}}$. Therefore the dispersion around both valleys can be unified in one formula $\epsilon_{\xi,\vect{k}}=\xi \pi_{\vect{k}}+\mu_{\vect{k}}$ (with $\xi=\pm$ for $K$ and $K'$ valleys). This enables us to rewrite \eqnref{eq:H0_supp} in the Majorana basis
\eq{H_0=\sum_{\vect{k},\xi=\pm}c_{\xi,\vect{k}}^\dagger (\xi \pi_{\vect{k}}+\mu_{\vect{k}})c_{\xi,\vect{k}}=\sum_{\vect{k}}c_{\vect{k}}^\dagger (\pi_{\vect{k}}\sigma^{30}+\mu_{\vect{k}}\sigma^{00})c_{\vect{k}}=\frac{1}{2}\sum_{\vect{k}}\chi_{-\vect{k}}^\intercal(\pi_{\vect{k}}\sigma^{300}+\mu_{\vect{k}}\sigma^{002})\chi_{\vect{k}},}
therefore the band electron action reads
\eq{S_0=\frac{1}{2}\sum_{k}\chi_{-k}^\intercal(-\ii\omega\sigma^{000}+\pi_{\vect{k}}\sigma^{300}+\mu_{\vect{k}}\sigma^{002})\chi_{k},}
where $k=(\ii\omega,\vect{k})$ is the frequency-momentum vector. We can write down the bare propagator
\eq{\label{eq:G0}G_0(k)\equiv-\langle\chi_{k}\chi_{-k}^\intercal\rangle=-(-\ii\omega\sigma^{000}+\pi_{\vect{k}}\sigma^{300}+\mu_{\vect{k}}\sigma^{002})^{-1}=\sum_{\vect{s}}\frac{P^\vect{s}}{\ii\omega-s_1\pi_{\vect{k}}-s_2\mu_{\vect{k}}},}
where the sign vector $\vect{s}=(s_1,s_2)$ (with $s_1,s_2=\pm$) labels different Majorana bands, and $P^\vect{s}$ is projection operator too the Majorana band
\eq{P^\vect{s}=\frac{\sigma^0+s_1\sigma^3}{2}\otimes\sigma^0\otimes\frac{\sigma^0+s_2\sigma^2}{2}.}

Substitute \eqnref{eq:G0} into \eqnref{eq:Pi_def}, the ladder kernel $\Pi$ can be evaluated as
\eqs{\Pi_{c'd'a'b'}(E)&=\sum_{\ii\omega}\sum_{k\in\scK_E}\sum_{\vect{s},\vect{s}'}\frac{P^{\vect{s}}_{c'b'}}{\ii\omega-s_1\pi_{\vect{k}}-s_2\mu_{\vect{k}}}\frac{P^{\vect{s}'}_{d'a'}}{-\ii\omega-s'_1\pi_{-\vect{k}}-s'_2\mu_{-\vect{k}}}\\
&=\sum_{k\in\scK_E}\sum_{\vect{s},\vect{s}'}-\frac{n_F(s_1\pi_\vect{k}+s_2\mu_\vect{k})-n_F(s'_1\pi_\vect{k}-s'_2\mu_\vect{k})}{(s_1-s'_1)\pi_\vect{k}+(s_2+s'_2)\mu_\vect{k}}P^{\vect{s}}_{c'b'}P^{\vect{s}'}_{d'a'}}
To further simplify, we con introduce the relative sign vector $\vect{\eta}=(\eta_1,\eta_2)$ such that $\vect{s}'=\vect{\eta}\odot\vect{s}$ (i.e.~$s'_i=\eta_is_i$ for $i=1,2$), then the ladder kernel can be classified into four different cases by $\vect{\eta}=++,+-,-+,--$
\eq{\label{eq:Pi_decom}\Pi(E)=\sum_{\vect{\eta}}\Pi^{\vect{\eta}}(E)Q^{\vect{\eta}},}
where $Q^{\vect{\eta}}$ is the two-particle projection operator,
\eq{\label{eq:Q}Q^{\vect{\eta}}=\sum_{\vect{s}}P^{\vect{s}}\otimes P^{\vect{\eta}\odot\vect{s}}}
 and the ladder diagram $\Pi^{\vect{\eta}}(E)$ is given by
\eq{\Pi^{\vect{\eta}}(E)=\sum_{k\in\scK_E}-\frac{n_F(\pi_\vect{k}+\mu_\vect{k})-n_F(\eta_1 \pi_\vect{k}-\eta_2 \mu_\vect{k})}{(1-\eta_1)\pi_\vect{k}+(1+\eta_2)\mu_\vect{k}},}
which can be enumerated in different $\vect{\eta}$-channels as follows:
\begin{itemize}
\item intra-valley particle-particle fluctuation ($\vect{\eta}=++$)
\eq{\Pi_{\vect{0}}^\text{pp}(E)\equiv\Pi^{++}(E)=\sum_{k\in\scK_E}-\frac{n_F(\pi_\vect{k}+\mu_\vect{k})-n_F(\pi_\vect{k}-\mu_\vect{k})}{2\mu_\vect{k}}\stackrel{T\to0}{=}\sum_{|\mu_\vect{k}|>E}\frac{\Theta(|\mu_\vect{k}|-|\pi_\vect{k}|)}{|\mu_\vect{k}|},}

\item intra-valley particle-hole fluctuation ($\vect{\eta}=+-$)
\eq{\Pi_{\vect{0}}^\text{ph}(E)\equiv\Pi^{+-}(E)=0,}

\item inter-valley particle-particle fluctuation ($\vect{\eta}=-+$)
\eq{\Pi_{\vect{Q}}^\text{pp}(E)\equiv\Pi^{-+}(E)=\sum_{k\in\scK_E}-\frac{n_F(\epsilon_\vect{k})-n_F(-\epsilon_\vect{k})}{2\epsilon_\vect{k}}\stackrel{T\to0}{=}\sum_{|\epsilon_\vect{k}|>E}\frac{1}{|\epsilon_\vect{k}|},}

\item inter-valley particle-hole fluctuation ($\vect{\eta}=--$)
\eq{\Pi_{\vect{Q}}^\text{ph}(E)\equiv\Pi^{--}(E)=\sum_{k\in\scK_E}-\frac{n_F(\pi_\vect{k}+\mu_\vect{k})-n_F(-\pi_\vect{k}+\mu_\vect{k})}{2\pi_\vect{k}}\stackrel{T\to0}{=}\sum_{|\pi_\vect{k}|>E}\frac{\Theta(|\pi_\vect{k}|-|\mu_\vect{k}|)}{|\pi_\vect{k}|}.}

\end{itemize}

Suppose the chemical potential is tuned to a van Hove singularity, and the density of state diverges logarithmically $2\nu_0\ln(\Lambda/\epsilon)$ as $\epsilon\to 0$. Then among all four $\Pi^\vect{\eta}(E)$, only $\Pi_{\vect{Q}}^\text{pp}(E)=\Pi^{-+}(E)\sim\nu_0\ln^2(\Lambda/E)$ will diverge double-logarithmically as $E\to 0$. Therefore instead of using the log energy cutoff $(-\ln E)$ as the RG scale, we would better use $y=\Pi_{\vect{Q}}^\text{pp}(E)$ as the RG scale. Reducing the energy scale $E$ corresponds to increasing the scale $y$. Introduce the following nesting parameters to track how the other ladder diagram scales with $y$,
\eq{\label{eq:nesting}d_{\vect{0}}^\text{pp}=\partial_y\Pi_{\vect{0}}^\text{pp}=\frac{\dd \Pi_{\vect{0}}^\text{pp}}{\dd \Pi_{\vect{Q}}^\text{pp}}, d_{\vect{Q}}^\text{ph}=\partial_y\Pi_{\vect{Q}}^\text{ph}=\frac{\dd \Pi_{\vect{Q}}^\text{ph}}{\dd \Pi_{\vect{Q}}^\text{pp}},}
then we have
\eq{\dd \Pi^{++}=d_{\vect{0}}^\text{pp}\dd y, \dd \Pi^{+-}=0, \dd \Pi^{-+}=\dd y, \dd \Pi^{--}=d_{\vect{Q}}^\text{ph}\dd y.}
Because $\Pi_{\vect{0}}^\text{pp}(E)$ and $\Pi_{\vect{Q}}^\text{ph}(E)$ do not diverge as fast as $\Pi_{\vect{Q}}^\text{pp}(E)$ as $E\to 0$, we expect the nesting parameters to fall in the regime of $d_{\vect{0}}^\text{pp},d_{\vect{Q}}^\text{ph} \in[0,1]$.

With the above parametrization of $\dd\Pi^\vect{\eta}$, we substitute \eqnref{eq:Pi_decom} into \eqnref{eq:dV}, and obtain 
\eq{\label{eq:dV2} \dd V_{abcd}=-\sum_{\vect{\eta}}V_{abc'd'}Q^\vect{\eta}_{c'd'a'b'}V_{a'b'cd}\;\dd\Pi^\vect{\eta}=-V_{abc'd'}(d_{\vect{0}}^\text{pp}Q^\vect{++}_{c'd'a'b'}+Q^\vect{-+}_{c'd'a'b'}+d_{\vect{Q}}^\text{ph}Q^\vect{--}_{c'd'a'b'})V_{a'b'cd}\;\dd y.}
Plug in the interaction vertex $V$ in \eqnref{eq:V} and the two-particle projection operator $Q^{\vect{eta}}$ in \eqnref{eq:Q} into \eqnref{eq:dV2}, we arrive at the RG equation
\eqs{
\tfrac{\dd}{\dd y}U&=-\tfrac{1}{2}d_{\vect{0}}^\text{pp}U^2,\\
\tfrac{\dd}{\dd y}V&=-\tfrac{1}{2}(1-d_{\vect{Q}}^\text{ph})(V^2+3J_H^2),\\
\tfrac{\dd}{\dd y}J_H&=-J_H((1+d_{\vect{Q}}^\text{ph})J_H+(1-d_{\vect{Q}}^\text{ph})V).}

We can show the RG equation is consistent with similar results in existing literature. In \refcite{Furukawa1998Truncation}, the one-loop RG of the following interaction was studied
\eq{\label{eq:Hint_g}H_\text{int}=\sum _{\alpha\neq\beta}(g_1 c_\alpha^\dagger c_\beta^\dagger c_\alpha c_\beta+g_2 c_\alpha^\dagger c_\beta^\dagger c_\beta c_\alpha+g_3 c_\alpha^\dagger c_\alpha^\dagger c_\beta c_\beta+)+\sum_{\alpha}g_4 c_\alpha^\dagger c_\alpha^\dagger c_\alpha c_\alpha,}
and the following RG equation was obtained
\eqs{\tfrac{\dd}{\dd y'}g_1&=-2 d_1 g_1(g_1-g_2)+2d_2g_1g_4-2d_3 g_1 g_2,\\
\tfrac{\dd}{\dd y'}g_2&=d_1(g_2^2+g_3^2)+2d_2(g_1-g_2)g_4-d_3(g_1^2+g_2^2),\\
\tfrac{\dd}{\dd y'}g_3&=-2g_3g_4-2d_1g_3(g_1-2g_2),\\
\tfrac{\dd}{\dd y'}g_4&=-(g_3^2+g_4^2)+d_2(g_1^2+2g_1g_2-g_2^2+g_4^2),\\
}
where the RG time is $y'=\Pi_{\vect{0}}^\text{pp}$ and the nesting parameters are defined as
\eq{d_1=\frac{\dd\Pi_{\vect{Q}}^\text{ph}}{\dd\Pi_{\vect{0}}^\text{pp}},d_2=\frac{\dd\Pi_{\vect{0}}^\text{ph}}{\dd\Pi_{\vect{0}}^\text{pp}},d_3=\frac{\dd\Pi_{\vect{Q}}^\text{pp}}{\dd\Pi_{\vect{0}}^\text{pp}}.}
We first need to change the RG time from $y'=\Pi_{\vect{0}}^\text{pp}$ to $y=\Pi_{\vect{Q}}^\text{pp}$ and rewrite the RG equation in terms of our nesting parameters in \eqnref{eq:nesting} (which amounts to replacing $\frac{\dd}{\dd y'}=(1/d_\vect{0}^\text{pp})\frac{\dd}{\dd y}$, $d_1=d_\vect{Q}^\text{ph}/d_\vect{0}^\text{pp}$, $d_2=0$, $d_3=1/d_\vect{0}^\text{pp}$),
\eqs{\tfrac{\dd}{\dd y}g_1&=-2 d_{\vect{Q}}^\text{ph} g_1(g_1-g_2)-2 g_1 g_2,\\
\tfrac{\dd}{\dd y}g_2&=d_{\vect{Q}}^\text{ph}(g_2^2+g_3^2)-(g_1^2+g_2^2),\\
\tfrac{\dd}{\dd y}g_3&=-2 d_\vect{0}^\text{pp} g_3g_4-2 d_{\vect{Q}}^\text{ph} g_3(g_1-2g_2),\\
\tfrac{\dd}{\dd y}g_4&=-d_\vect{0}^\text{pp} (g_3^2+g_4^2).\\
}
Then by comparing with \eqnref{eq:H_int_supp}, we identify the couplings in \eqnref{eq:Hint_g} as
\eq{g_1=J_H,g_2=\frac{1}{2}(V+J_H), g_3=0,g_4=\frac{1}{2}U,}
then the RG equation becomes
\eqs{\tfrac{\dd}{\dd y}J_H&=- d_{\vect{Q}}^\text{ph} J_H(J_H-V)-J_H (J_H+V),\\
\tfrac{\dd}{\dd y}\tfrac{1}{2}(V+J_H)&=d_{\vect{Q}}^\text{ph}\tfrac{1}{4}(V+J_H)^2-(J_H^2+\tfrac{1}{4}(V+J_H)^2),\\
\tfrac{\dd}{\dd y}(\tfrac{1}{2}U)&=-d_\vect{0}^\text{pp} (\tfrac{1}{4}U^2).\\
}
After some simple algebra, it is straightforward to show that the RG equation is consistent with our derivation.

\subsection{Other Form of Hund's Interaction}\label{app:Hunds}

\begin{figure}[htbp]
\begin{center}
\includegraphics[width=0.6\columnwidth]{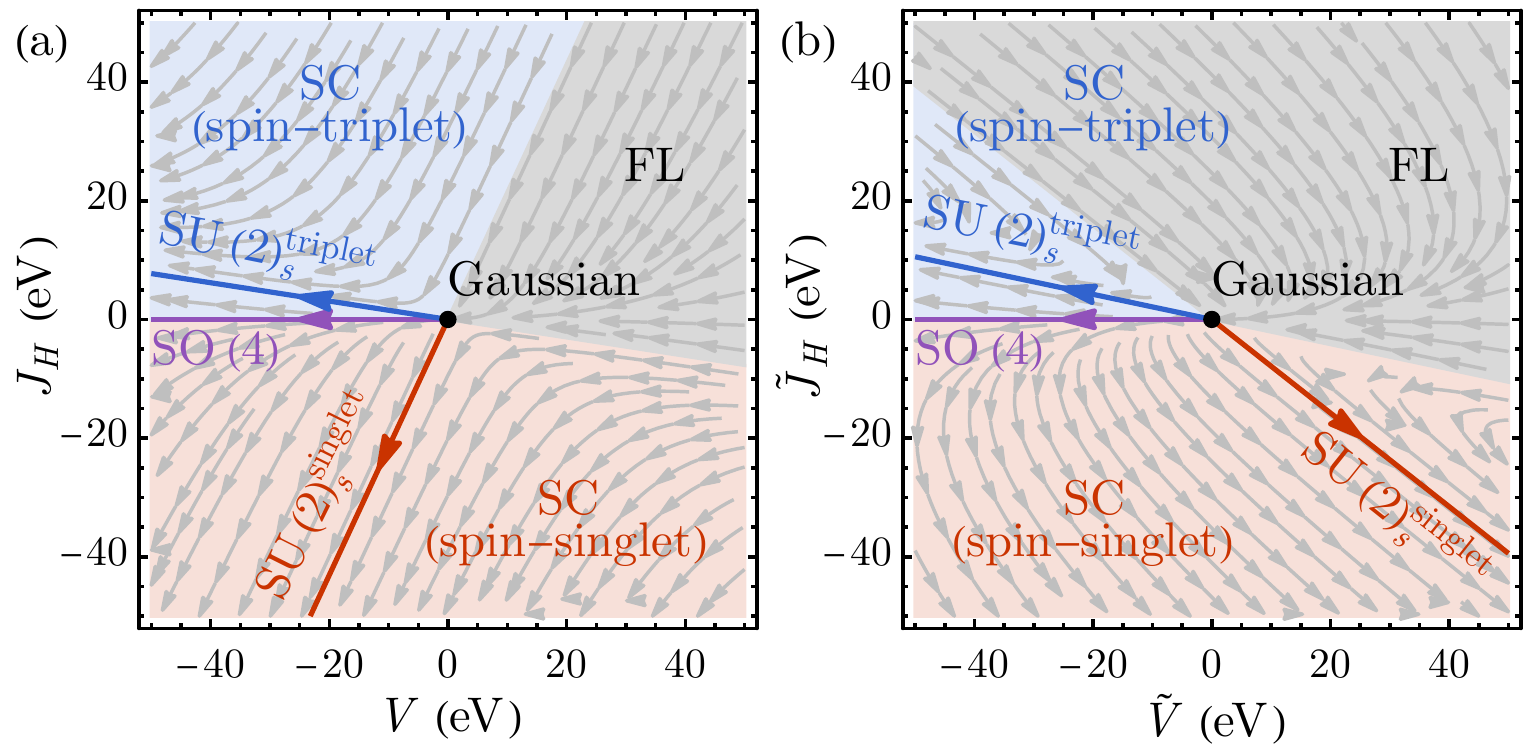}
\caption{(a) RG flow in $(V,J_H)$ with spin Hund's interaction following \eqnref{eq:RG1}. (b) RG flow in $(\tilde{V},\tilde{J}_H)$ with IVC Hund's interaction following \eqnref{eq:RG2}. Both RG flow diagrams are calculated for a nesting parameter $d_\vect{Q}^{ph}=0.5$. Note the red (blue) regions correspond to spin singlet (triplet) superconductor, while the grey is the metal (Fermi liquid).}
\label{fig: RG2}
\end{center}
\end{figure}

 Another form \cite{You2019Superconductivity} of  Hund's interaction in terms of IVC operators $\vect{I}=c_{K}^\dagger \vect{\sigma} c_{K'}$, (\refcite{Chatterjee2021Inter-valley} points out that this has a better justified microscopic origin in the present problem). In terms of the IVC Hund's coupling, the total interaction will be reparametrized as
\eq{\label{eq:H_int2}H_\text{int}=\int\dd^2\vect{x}\;\frac{U}{2}(n_{K}^2+n_{K'}^2)+\tilde{V}n_{K}n_{K'}-\tilde{J}_H\vect{I}^\dagger\cdot\vect{I}.}
We have assume the limit that the interaction is very local, such that there is no momentum dependence of the coupling coefficients. In this limit, we can use the following operator identity
\eq{\vect{I}^\dagger\cdot\vect{I}=-\frac{3}{2}n_K n_{K'}+\frac{1}{2}\vect{S}_K\cdot\vect{S}_{K'}}
to connect \eqnref{eq:H_int2} and \eqnref{eq:H_int_supp}, and establish the following linear relations between couplings $(V,J_H)$ and $(\tilde{V},\tilde{J}_H)$,
\eq{\mat{V\\J_H}=\mat{1&3/2\\0&1/2}\mat{\tilde{V}\\\tilde{J}_H}.}
Therefore the RG equations
\eqs{\label{eq:RG1}
\tfrac{\dd}{\dd y}U&=-\tfrac{1}{2}d_{\vect{0}}^\text{pp}U^2,\\
\tfrac{\dd}{\dd y}V&=-\tfrac{1}{2}(1-d_{\vect{Q}}^\text{ph})(V^2+3J_H^2),\\
\tfrac{\dd}{\dd y}J_H&=-J_H((1+d_{\vect{Q}}^\text{ph})J_H+(1-d_{\vect{Q}}^\text{ph})V),\\}
can also be written as
\eqs{\label{eq:RG2}
\tfrac{\dd}{\dd y}U&=-\tfrac{1}{2}d_{\vect{0}}^\text{pp}U^2,\\
\tfrac{\dd}{\dd y}\tilde{V}&=-\tfrac{1}{2}(1-d_{\vect{Q}}^\text{ph})\tilde{V}^2+\tfrac{3}{2}\tilde{J}_H^2,\\
\tfrac{\dd}{\dd y}\tilde{J}_H&=-\tilde{J}_H((2-d_{\vect{Q}}^\text{ph})\tilde{J}_H+(1-d_{\vect{Q}}^\text{ph})\tilde{V}),\\}
The RG equation for $U$ is not affected by the reparameterization, and is decoupled. The RG flow in the $(V,J_H)$ or $(\tilde{V},\tilde{J}_H)$ planes are compared in \figref{fig: RG2}. 

\subsection{RG Fixed Rays and Oder Parameter Scaling}\label{app:scaling}

We will focus on the $(V,J_H)$ basis, and discuss the RG fixed rays of the flow equation \eqnref{eq:RG1}. By solving the RG equation, the asymptotic behavior near $y\to y_c$ is expected to be
\eq{\label{eq:RG_sol}V=\frac{\gamma_V}{y_c-y}, J_H = \frac{\gamma_{J_H}}{y_c-y}.}
Substitute \eqnref{eq:RG_sol} to \eqnref{eq:RG1}, one can determine the coefficients $\gamma_V$ and $\gamma_{J_H}$ by the following algebraic equations
\eqs{\gamma_V&=\tfrac{1}{2}(1-d_\vect{Q}^\text{ph})(\gamma_V^2+3 \gamma_{J_H}^2),\\
\gamma_{J_H}&=-\gamma_{J_H}((1+d_\vect{Q}^\text{ph})\gamma_{J_H}+(1-d_\vect{Q}^\text{ph})\gamma_V).}
The solutions are:
\begin{itemize}
\item Gaussian fixed point
\eq{\gamma_V=\gamma_J=0;}
\item $\SO(4)$ fixed ray
\eq{\gamma_V=-\frac{2}{1-d_\vect{Q}^\text{ph}}, \gamma_{J_H}=0;}
\item $\SU(2)_\text{s}^\text{singlet}$ fixed ray
\eq{\gamma_V=-\frac{1}{1-d_\vect{Q}^\text{ph}}\left(1-\frac{1+d_\vect{Q}^\text{ph}}{2\sqrt{1-d_\vect{Q}^\text{ph}+(d_\vect{Q}^\text{ph})^2}}\right), \gamma_{J_H}=-\frac{1}{2\sqrt{1-d_\vect{Q}^\text{ph}+(d_\vect{Q}^\text{ph})^2}};}
\item $\SU(2)_\text{s}^\text{triplet}$ fixed ray
\eq{\gamma_V=-\frac{1}{1-d_\vect{Q}^\text{ph}}\left(1+\frac{1+d_\vect{Q}^\text{ph}}{2\sqrt{1-d_\vect{Q}^\text{ph}+(d_\vect{Q}^\text{ph})^2}}\right), \gamma_{J_H}=\frac{1}{2\sqrt{1-d_\vect{Q}^\text{ph}+(d_\vect{Q}^\text{ph})^2}}.}
\end{itemize}

To study the scaling of fermion bilinear order parameters along the RG flow, one can consider deforming the Hamiltonian with infinitesimal fermion bilinear terms $H=H_0+\delta H$,
\eq{\delta H=\int\dd^2 x\sum_{a,b}X_{ab}\chi_a\chi_b,}
where $X$ is an antisymmetric imaginary matrix representing the fermion bilinear order. The correction of $X$ is given by
$X_{ab}(E)=X_{ab}-V_{abc'd'}\Pi_{c'd'a'b'}(E)X_{a'b'}+\cdots$,
which leads to the RG equation
\eq{\dd X_{ab}=-V_{abc'd'}\dd\Pi_{c'd'a'b'} X_{a'b'}=-V_{abc'd'}(d_{\vect{0}}^\text{pp}Q^\vect{++}_{c'd'a'b'}+Q^\vect{-+}_{c'd'a'b'}+d_{\vect{Q}}^\text{ph}Q^\vect{--}_{c'd'a'b'}) X_{a'b'}.}

Along the eigen direction (labeled by $\Phi$) of the RG flow, we have
\eq{\tfrac{\dd}{\dd y}X_\Phi=V_\Phi X_\Phi,}
where $V_\Phi$ denotes the eigen value (which depends on $V_{abcd}$ linearly). Near the critical point $y\to y_c$, $V_\Phi$ diverges as $\gamma_\Phi/(y_c-y)$, then the RG equation implies
\eq{\frac{\dd}{\dd y} X_\Phi=\frac{\gamma_\Phi}{y_c-y}X_\Phi,}
whose solution is
\eq{X_\Phi(y)\propto(y_c-y)^{-\gamma_\Phi}.}
Then the susceptibility diverges near $y_c$ as
\eq{\chi_\Phi(y)=\frac{X_\alpha(y)}{X_\Phi(0)}\sim(y_c-y)^{-\gamma_\Phi}.}
Therefore, a large positive exponent $\gamma_\Phi$ indicates a stronger instability in the eigen channel of the fermion bilinear ordering. In the following, we will apply this general framework to different bilinear orders.

\begin{itemize}
\item Inter-Valley superconductivity $\Delta^\mu=c_{K}^\intercal\ii\sigma^2\sigma^\mu c_{K'}$. The RG equation reads
\eqs{\tfrac{\dd}{\dd y}\Delta^0&=-2(3J_H+V)\Delta^0,\\
\tfrac{\dd}{\dd y}\vect{\Delta}&=2(J_H-V)\vect{\Delta},\\}
therefore $\gamma_{\Delta^0}=-2(3\gamma_{J_H}+\gamma_{V})$ and $\gamma_{\vect{\Delta}}=2(\gamma_{J_H}-\gamma_{V})$.

\item Inter-Valley coherence $I^\mu=c_{K}^\dagger\sigma^\mu c_{K'}$. The RG equation reads
\eqs{\tfrac{\dd}{\dd y}I^0&=-2d_\vect{Q}^\text{ph}(3J_H-V)I^0,\\
\tfrac{\dd}{\dd y}\vect{I}&=2d_\vect{Q}^\text{ph}(J_H+V)\vect{I},\\}
therefore $\gamma_{\Delta^0}=-2d_\vect{Q}^\text{ph}(3\gamma_{J_H}-\gamma_{V})$ and $\gamma_{\vect{\Delta}}=2d_\vect{Q}^\text{ph}(\gamma_{J_H}+\gamma_{V})$.

\item Intra-Valley superconductivity $\Delta_\xi=c_{\xi}^\intercal \ii\sigma^2 c_{\xi}$. The RG equation reads
\eqs{\tfrac{\dd}{\dd y}\Delta_\xi&=-2d_\vect{0}^\text{pp}U\Delta_\xi,}
therefore $\gamma_{\Delta_\xi}=-2d_\vect{0}^\text{pp}\gamma_U$.

\item Valley charge density $n_v=n_{K}-n_{K'}$. The RG equation is trivial $\tfrac{\dd}{\dd y}n_v=0$, so $\gamma_{n_v}=0$.

\item Valley charge density $\vect{S}_{\xi}=c_{\xi}^\dagger \vect{\sigma} c_{\xi}$. The RG equation is trivial $\tfrac{\dd}{\dd y}\vect{S}_{\xi}=0$, so $\gamma_{\vect{S}_{\xi}}=0$.
\end{itemize}
\end{document}